\documentclass[onecolumn,superscriptaddress,nofootinbib,notitlepage,longbibliography,aps]{revtex4-1}
\pdfoutput=1

\usepackage{graphicx}
\usepackage{amsmath}
\usepackage{amssymb}
\usepackage{amsfonts}
\usepackage[dvipsnames]{xcolor}
\usepackage[linktoc=none]{hyperref}
\usepackage[utf8]{inputenc}
\hypersetup{colorlinks=true,linkcolor=blue,citecolor=teal,filecolor=magenta, urlcolor=cyan}

\newcommand{\INFN}{INFN - Sezione di Napoli, Complesso Univ. Monte S. Angelo, I-80126 Napoli, Italy}
\newcommand{\UNINA}{Dipartimento di Fisica ``Ettore Pancini'', Università degli studi di Napoli ``Federico II'', Complesso Univ. Monte S. Angelo, I-80126 Napoli, Italy}
\newcommand{\SSM}{Scuola Superiore Meridionale, Università degli studi di Napoli ``Federico II'', Largo San Marcellino 10, 80138 Napoli, Italy}

\begin{document}

\title{Sensitivity of KM3NeT to Violation of Equivalence Principle}

\author{Marco Chianese}
\affiliation{\UNINA}
\affiliation{\INFN}
\author{Damiano F.G. Fiorillo}
\email{dfgfiorillo@na.infn.it}
\affiliation{\UNINA}
\affiliation{\INFN}
\author{Gianpiero Mangano}
\affiliation{\UNINA}
\affiliation{\INFN}
\author{Gennaro Miele}
\affiliation{\UNINA}
\affiliation{\INFN}
\affiliation{\SSM}
\author{Stefano Morisi}
\affiliation{\INFN}
\affiliation{\SSM}
\author{Ofelia Pisanti}
\affiliation{\INFN}
\affiliation{\SSM}

\date{\today}
\begin{abstract}
The symmetry of the theory of relativity under diffeomorphisms strongly depends on the equivalence principle. Violation of Equivalence Principle (VEP) can be tested by looking for deviations from the standard framework of neutrino oscillations. In recent works, it has been shown that strong constraints on the VEP parameter space can be placed by means of the atmospheric neutrinos observed by the IceCube neutrino telescope. In this paper, we focus on the KM3NeT neutrino telescope and perform a forecast analysis to assess its capacity to probe VEP. Most importantly, we examine the crucial role played by systematic uncertainties affecting the neutrino observations. We find that KM3NeT will constrain VEP parameters times the local gravitational potential at the level of $10^{-27}$. Due to the systematic-dominated regime, independent analyses from different neutrino telescopes are fundamental for robustly testing the equivalence principle.
\end{abstract}

\maketitle

\section{Introduction}
The equivalence principle is a fundamental assumption of general relativity, according to which all test points follow identical trajectories in the same gravitational field. {By virtue of this principle, the effects of gravity can be locally described by a non-inertial system of coordinates---this physical statement allows to describe the theory of gravity in a form which is invariant under general transformations of coordinates.} The~weak equivalence principle implies that all particles couple to the gravitational field with the same coupling strength, parameterized by the Newton constant $G_N$. Observing a Violation of the Equivalence Principle (VEP) would be a shift in our perspective of gravity and would be an important piece of information in the construction of a complete theory of gravity. A~variety of experiments are suitable for testing VEP, including torsion balance experiments~\cite{Wagner:2012ui}, motion of bodies in the solar system~\cite{Overduin:2013soa}, spectroscopy of atomic levels~\cite{Hohensee:2013cya} and pulsars~\cite{Damour:1991rq,Horvat:1998st,Barkovich:2001rp}. 

In this paper, we focus on the possibility of probing VEP with high-energy neutrinos. VEP induces an energy shift between different combination of flavor eigenstates of neutrinos propagating in a gravitational potential. This energy shift leads to non-standard oscillations which are in principle distinguishable from the standard ones because of the different energy dependence. Interestingly, VEP was originally proposed to explain the solar neutrino problem~\cite{Pantaleone:1992ha,Butler:1993wi,Bahcall:1994zw,Halprin:1995vg,Mureika:1995ap,Mureika:1996de,Mureika:1996ud,Mansour:1998nb,Gago:1999hi,Casini:1999kt,Majumdar:2000sd}. Neutrino oscillations were later proposed as a powerful way to constrain VEP employing the non-observation of a deviation from the standard oscillations~\cite{Gasperini:1988zf,Gasperini:1989rt,Minakata:1994kt,Valdiviesso:2011zz,Foot:1997kk,Foot:1998pv,Fogli:1999fs,GonzalezGarcia:2004it,GonzalezGarcia:2005xw,Battistoni:2005gy,Morgan:2007dj,Abbasi:2009nfa,Pakvasa:1988gd,Guzzo:2001vn,Minakata:1996nd,Iida:1992vh,Mann:1995nw,Esmaili:2014ota,Diaz:2020aax}. We emphasize that the constraints obtained on VEP through this methodology are hardly comparable with the constraints obtained from purely gravitational experiments. On~the one hand, non-standard oscillations induced by VEP can probe relative differences in the gravitational constant of the order of $10^{-22}$--$10^{-23}$, whereas experiments based on gravitational effects only reach at most a sensitivity of $10^{-13}$~\cite{Tino:2020nla}. On~the other hand, however, non-standard oscillations can only probe differences in the gravitational coupling among different flavor states of neutrinos, and~therefore refer only to a specific model of~VEP.

The effects of VEP are more important at high energies, and~therefore an ideal class of experiments for testing it are neutrino telescopes~\cite{Aslanides:1999vq,Aartsen:2014oha,Barwick:2007vba,Meures:2012fka,Martineau-Huynh:2015hae, Schulz:2009zz,Blaufuss:2015muc,Collaboration:2011nsa,Belolaptikov:1997ry,Riccobene:2017fpr,Kappes:2007ci,Goldschmidt:2001qd,Ahrens:2002dv}. In~recent works~\cite{Fiorillo:2020gsb, Esmaili:2014ota}, the~atmospheric neutrinos detected by IceCube have been used to constrain the parameter space of VEP. However, the~related question of how these constraints could improve with increasing observation time and with the future neutrino telescopes under construction~\cite{Adrian-Martinez:2016fdl,Aartsen:2020fgd} has not been explored. This question is even more important because, as~emphasized in Ref.~\cite{Fiorillo:2020gsb}, the~constraints obtained are strongly dependent on the systematic error affecting the~data.

Along this direction, in~this work we explore the potentialities of KM3NeT~\cite{Adrian-Martinez:2016fdl} to constrain VEP. We perform a forecast analysis of the atmospheric muon neutrinos that will be observed by KM3NeT and obtain the projected constraints after 10 years of observation. We assess the role of the systematic error by considering it as a free parameter, showing that indeed the resulting constraint can significantly depend on it. The~structure of the work is as follows: in Section~\ref{sec:oscillations} we describe the general framework of neutrino oscillations in the presence of VEP, and~the methods we adopt to describe them; in Section~\ref{sec:atmoic} we describe the forecast analysis performed in order to estimate the projected constraints at KM3NeT. Finally, in~Section~\ref{sec:concl} we draw our~conclusions.

\section{Neutrino Oscillations in the Presence Of~VEP} \label{sec:oscillations}

In this section we introduce the formalism adopted for describing the non-standard oscillations induced by VEP. Throughout this work we use units in which $\hbar=c=1$. Let us study the propagation of neutrinos in matter, with~a number density $N_e(\mathbf{r})$ of electrons, in~the presence of a gravitational potential $\phi(\mathbf{r})$; we denote by $c_\alpha$ the amplitudes of the $\alpha=e,\;\mu,\;\tau$ flavor. If~the equivalence principle is violated, three combinations of the flavor eigenstates are selected which couple with three different strengths $\gamma_a$ (the index $a$ runs from 1 to 3) to the gravitational potential. The~amplitudes for these new states, which are eigenstates of the gravitational interaction, are defined by $c_a=\sum_\alpha \tilde{U}_{\alpha a} c_\alpha$, where $\tilde{U}$ is a unitary matrix. As~discussed in Ref.~\cite{Fiorillo:2020gsb}, there is a considerable model dependence in the choice of $\tilde{U}$, which can lead to observable effects even for astrophysical neutrinos. However, since for this work we are mainly interested in atmospheric neutrinos, we will stick to the more common choice that $\tilde{U}=U$, where $U$ is the PMNS matrix. In~the following, for~the neutrino conventional oscillation parameters we adopt the best-fit values from Ref.~\cite{Zyla:2020zbs} and consider the scenario of direct neutrino mass~ordering.

Following Ref.~\cite{Fiorillo:2020gsb} (see also Ref.~\cite{Valdiviesso:2011zz}), the~equations describing the neutrino propagation along the path with length element $\mathrm{d}l$ are:
\begin{eqnarray}\label{eq:propavep}
i\frac{\mathrm{d}c_\alpha}{\mathrm{d}l}=\sum_{j,\beta} U_{\beta j} U^*_{\alpha j} \frac{\delta m_j^2}{2E} c_\beta + V(\mathbf{r}) \delta_{\alpha e} c_e
 +2E\phi\sum_{a,\beta} U_{\beta a} U^*_{\alpha a} \gamma_a c_\beta\,.
\end{eqnarray}

Here, $\delta m_j^2$ are the neutrino mass splittings and $E$ is the neutrino energy. Furthermore, $V(\mathbf{r})=\sqrt{2} G_F N_e(\mathbf{r})$ (being $G_F$ the Fermi constant) is the matter potential originating from the forward scattering of electron neutrinos off electrons. The~three couplings $\gamma_a$ are independent of one another. However, only their differences $\gamma_{21}=\gamma_2-\gamma_1$ and $\gamma_{31}=\gamma_3-\gamma_1$ can lead to physical effects. In~fact, subtracting from Equation~\eqref{eq:propavep}, the term $2E\gamma_1 c_\alpha$ causes only a global phase factor in front of all the amplitudes. In~the following, we will express all of our results in terms of the two differences $\gamma_{21}$ and $\gamma_{31}$ only.

The numerical solution of Equation~\eqref{eq:propavep} for neutrinos propagating through the Earth, assuming the preliminary reference Earth model (PREM)~\cite{Dziewonski:1981xy} for the electron number density, is described in detail in Ref.~\cite{Fiorillo:2020gsb}. For~this work, we adopt an approximate procedure which has however been found to provide reliable results. This procedure is similar to the one adopted for the similar case of Lorentz invariance violation in Ref.~\cite{Aartsen:2017ibm}. In~particular, the~equations can be simplified by noticing that the neutrino propagation is dominated by the matter term proportional to $V(\mathbf{r})$, which is of the order of $10^{-22}$ GeV, where we have used $N_e\sim 5\times 10^{24}$ cm$^{-3}$. Even for $\gamma_{21,31}\phi\sim 10^{-26}$, which are the largest values we consider and which are already excluded by the current constraints, the~effects due to VEP start dominating the matter effects only at energies greater than 10 TeV. Even though, as~we will discuss in the next section, our analysis is extended to energies of 100 TeV, only a small fraction of the events we analyze comes from this region. For~this reason, the~VEP effects can be treated as a small perturbation. In~this scenario, the~matter effect pins the electron amplitude $c_e$ to its initial value, and~only muon and tau neutrinos participate in oscillations. The~problem therefore is reduced to a simple two-flavor oscillation in vacuum, where the effective two-dimensional oscillation mass matrix is obtained by projecting Equation~\eqref{eq:propavep} in the $\mu$--$\tau$ basis. 

\section{Forecast~Analysis} \label{sec:atmoic}

Our aim in this section is to determine the projected constraints that KM3NeT will be able to place on the VEP model discussed in the previous section. As~in Ref.~\cite{Fiorillo:2020gsb}, we focus on upgoing atmospheric muon neutrinos with energies from 100~GeV up to 100~TeV and zenith angles larger than $90^\circ$. There are three reasons for this choice: first of all, it allows us to reject the contamination from atmospheric muons, which are attenuated in their propagation through the Earth; second, upgoing neutrinos travel a much larger distance compared to downgoing ones and, therefore, they are more sensitive to VEP oscillations; third, thanks to their track signature in the detector, muon neutrinos have a better angular resolution compared to the other flavors. In~the standard scenario (i.e., no VEP), neutrinos would be practically free from oscillations, since the wavelength is larger than the Earth's diameter: oscillations are only present at energies near 100 GeV. Let us discuss how this scenario changes in the presence of~VEP.

\begin{figure}[t!]
    \centering
    \includegraphics[width=6.5cm]{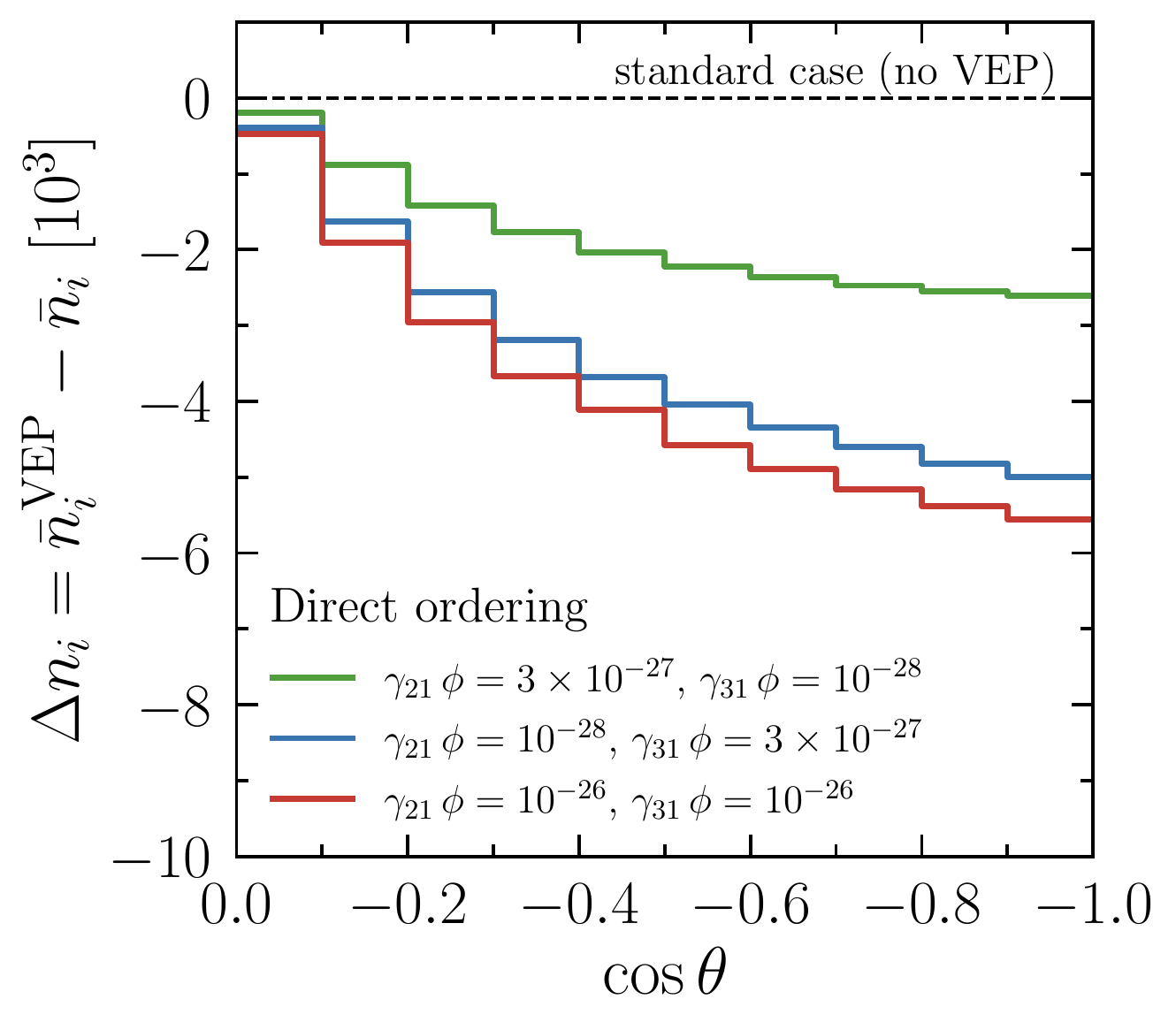}
    \hspace{0.2cm}
    \includegraphics[width=6.5cm]{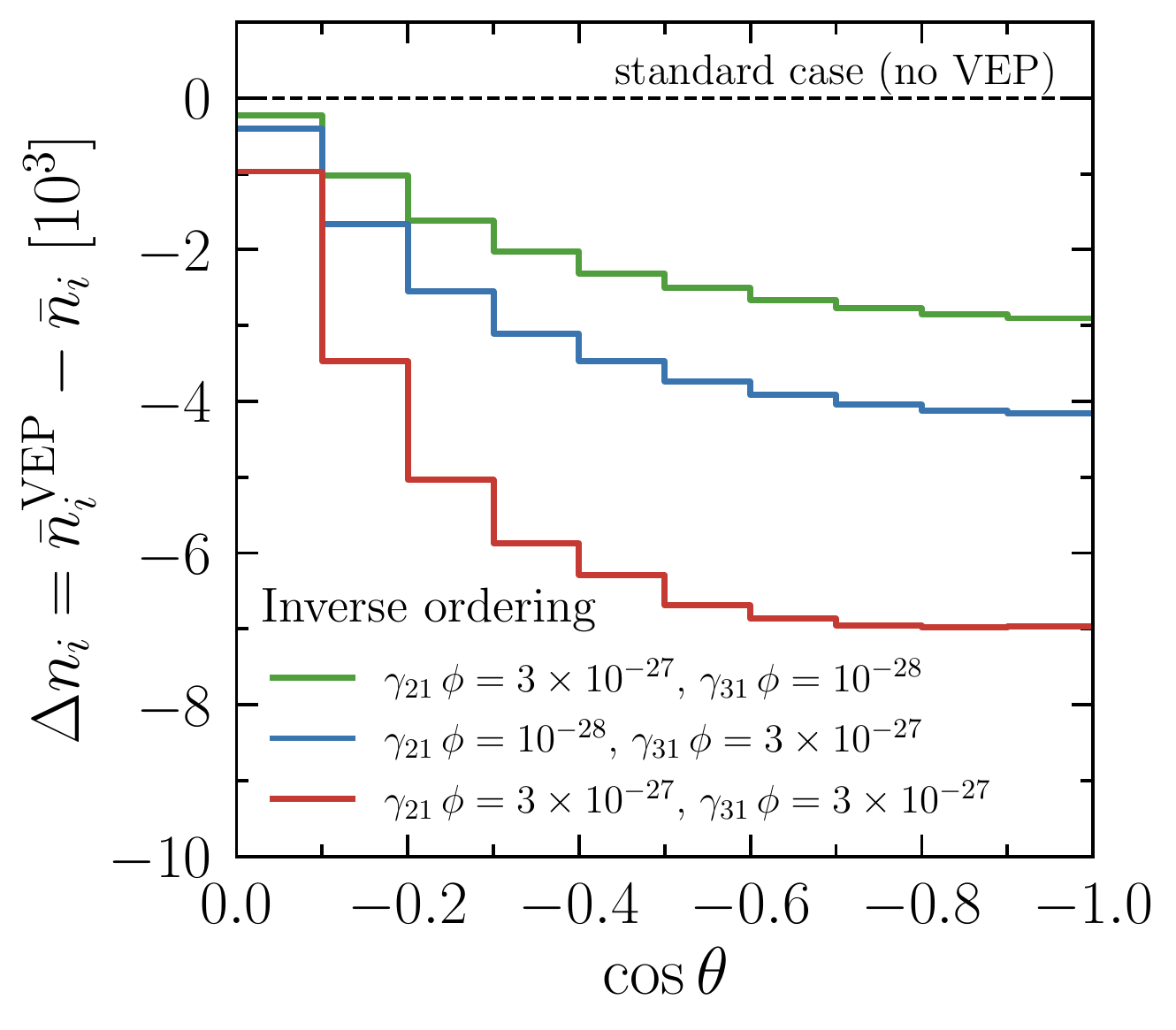}
    \caption{Variation as a function of the zenith angle on the expected KM3NeT number of neutrino events in 10 years of observation with respect to the standard oscillation scenario for different values of VEP parameters. Left (right) panel corresponds to the case of direct (inverse) ordering of VEP parameters. The~effect of VEP is a deficit of vertical upgoing neutrinos. All the uncertainties affecting the expected events are here~neglected.}
    \label{fig:vep_events}
\end{figure}

Neutrinos passing through the Earth are subjected to an average gravitational potential originating from the Great Attractor~\cite{Kenyon:1990ve} which, in~dimensionless units, is of the order of $\left|\phi\right|\sim10^{-5}$. This potential is so large as to cover completely the Earth potential, which is of the order of $10^{-9}$ in the same units. In~the VEP model, therefore, neutrinos propagate through the essentially constant gravitational potential $\phi$ for a distance $d \sim - 2 R \cos\theta$, where $R$ is the Earth radius and $\theta$ is the zenith angle (since $\theta>90^\circ$, $\cos\theta < 0$). During~their propagation, neutrinos with energies larger than $E\geq 1/\gamma\left|\phi\right| R$, where $\gamma$ is the largest between $\gamma_{21}$ and $\gamma_{31}$, are subject to VEP oscillations. Electron neutrinos are not substantially involved in the oscillation because of the forward scattering on nuclei. The~main effect is therefore a depletion of the atmospheric muon neutrino flux, since some of the muon neutrinos oscillate into the tau flavor. The~depletion is greater at zenith angles close to $180^\circ$, i.e.,~for nearly vertical neutrinos, because~of the larger distance traversed. Hence, the signature that can be constrained is the anisotropic deficit of events introduced by VEP. This depletion is depicted in Figure~\ref{fig:vep_events}, where we show the difference induced by VEP in the number of events compared to the standard case in ten bins of the zenith angle. Since the difference between the number of events with and without VEP is much smaller than the total number of events, we do not show the latter separately. The~VEP parameters are fixed to benchmark values. For~all choices of parameters, the~depletion is greater at zenith angles close to $180^\circ$. The~methods by which we have computed the data shown in Figure~\ref{fig:vep_events} are detailed~below.

In Ref.~\cite{Fiorillo:2020gsb} we have constrained VEP using the observed IceCube data. Since here we want to determine projected constraints for KM3NeT, we perform a Monte Carlo generation of mock data samples under the no-VEP hypothesis. As~in our previous work, we adopt ten equally spaced bins in zenith angle. In~each bin, the~expected number of event is
\begin{equation}\label{eq:expected}
    \bar{n}_i=2\pi T_\text{obs} \int \mathrm{d}E \int_{\cos\theta_i}^{\cos\theta_{i+1}} \mathrm{d}\cos\theta\,\frac{\mathrm{d}\Phi_\text{atm}}{\mathrm{d}E\mathrm{d}\Omega} A_\text{eff}(E) \,.
\end{equation}

Here, $T_\text{obs}$ is the observation time (specified below), $\mathrm{d}\Phi_\text{atm}/\mathrm{d}E\mathrm{d}\Omega$ is the atmospheric muon neutrino flux, and~$A_\text{eff}(E)$ is the effective area of KM3NeT for muon neutrino detection~\cite{Adrian-Martinez:2016fdl}. For~the atmospheric muon neutrino flux we adopt the model discussed in Ref.~\cite{Honda:2006qj} with a normalization larger by $15\%$. This assumption has resulted in being in better agreement with the real IceCube data analyzed in our previous work. However, we emphasize that this $15\%$ increase in the normalization is well contained in the estimated theoretical uncertainty of $24\%$ on the flux normalization (see Refs.~\cite{Honda:2006qj, Esmaili:2014ota, Fiorillo:2020gsb} and below). Concerning the KM3NeT effective area, only its dependence on the neutrino energy is publicly available, while information on its angular dependence is not provided. Since our analysis critically depends on the angular distribution of events, we note that this information could sensibly influence our~results.

In the Monte Carlo generation of mock data samples, a~crucial element to be taken into account is the systematic error affecting the experimental measurement (hereafter denoted as $\bar{f}_\mathrm{sys}$). As~shown in our previous work~\cite{Fiorillo:2020gsb}, it is not realistic to assume that data are subject to statistical Poisson fluctuations only. In~fact, in~the absence of the systematic error ($\bar{f}_\mathrm{sys}=0$), the~IceCube data result in not being in good agreement with the model of atmospheric fluxes. In~our previous work, following the method of Ref.~\cite{Esmaili:2014ota}, we estimated the systematic error from the data such that an agreement with the theoretical flux model is achieved at the $2\sigma$ level. The~typical systematic errors we found for different IceCube datasets range from $1\%$ to $4\%$. For~this reason, in~generating mock data samples for KM3NeT we cannot neglect a possible systematic uncertainty. Hence, we produce the data according to a normal distribution for each angular bin centered around the expected value $\bar{n}_i$ with standard deviation $\sigma^2_i=\bar{n}_i+\bar{f}_\mathrm{sys}^2\bar{n}_i$. We adopt two different benchmark choices for the generation: $\bar{f}_\mathrm{sys} = 0.04$ and $\bar{f}_\mathrm{sys} = 0.01$.
\begin{figure}[t!]
    \centering
    \includegraphics[width=6.5cm]{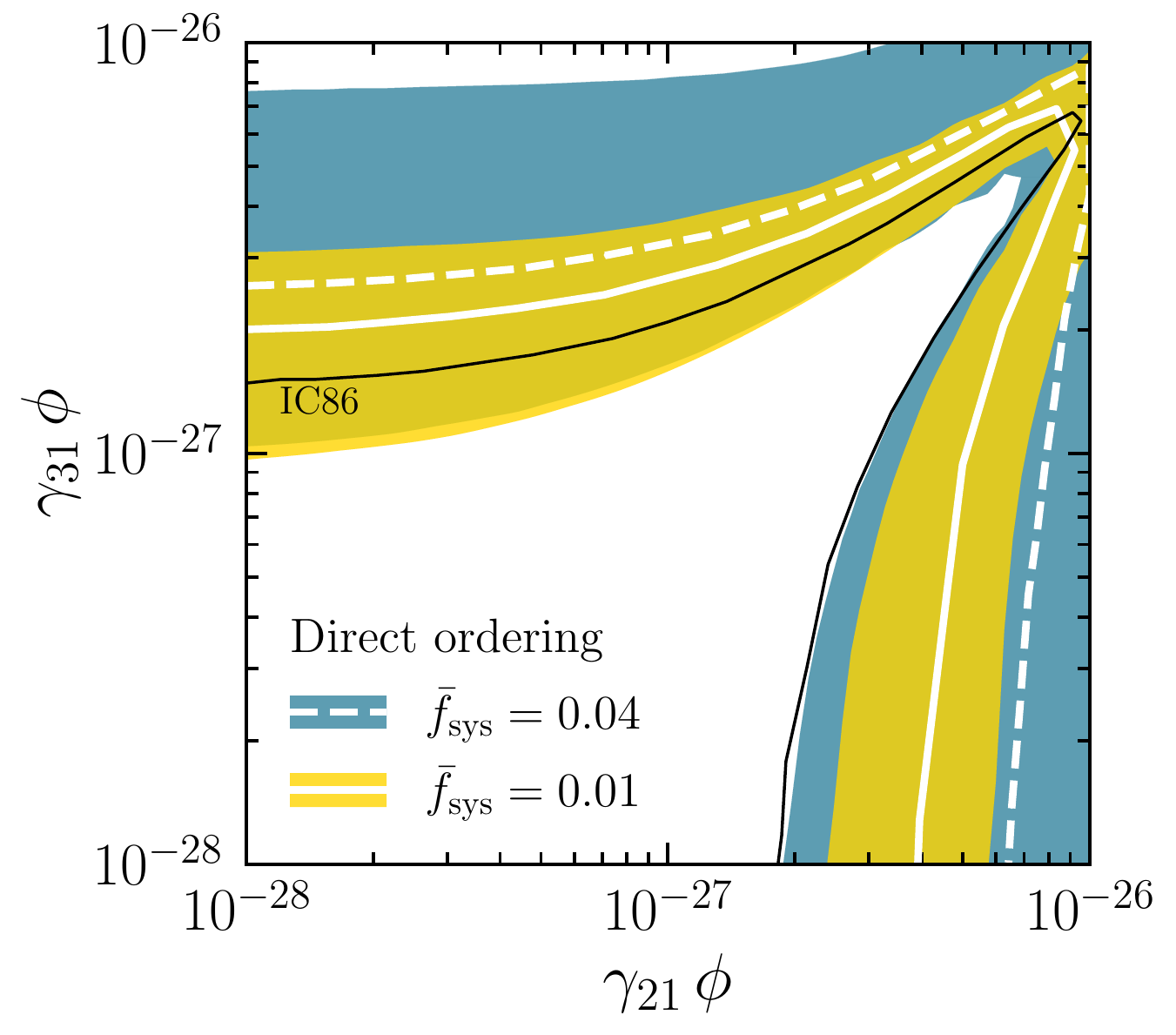}
    \hspace{0.2cm}
    \includegraphics[width=6.5cm]{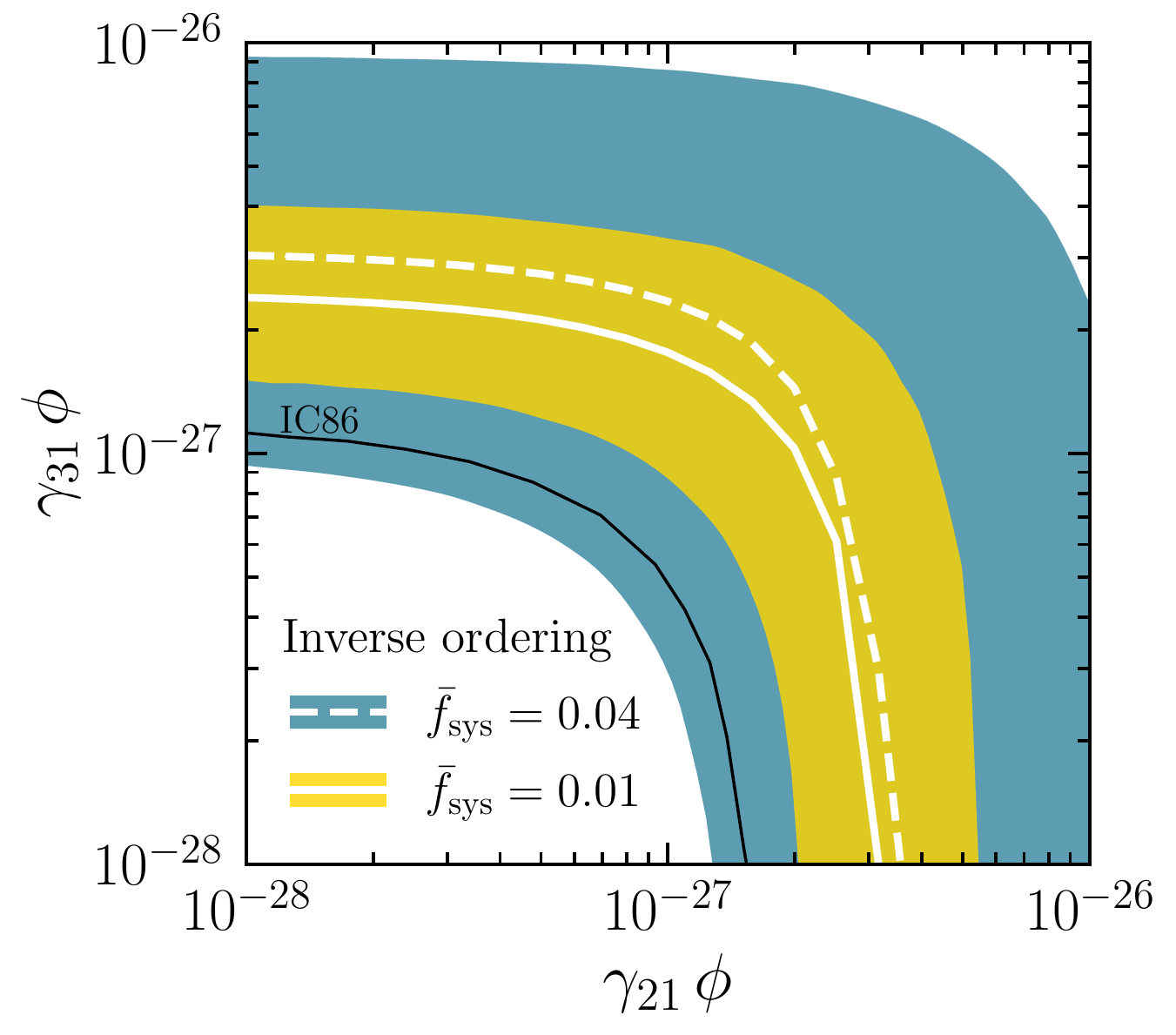}
    \caption{KM3NeT-forecasted 10-year constraints at 90\% CL on the VEP parameters for direct ordering ($\gamma_{21}\times\gamma_{31}>0$) and inverse ordering ($\gamma_{21}\times\gamma_{31}<0$). The~bands cover all the Monte Carlo simulations of mock data samples. The~blueish and yellow colors correspond to a systematic error of 0.01 and 0.04 in the data generation, respectively. The~solid and dashed white lines are the mean constrains for the two values of the systematic error. The~thin black lines show the IceCube limits obtained by means of a similar analysis~\cite{Fiorillo:2020gsb}.}
    \label{fig:exclusioncontours10years}
\end{figure}

For each data sample $n_i$ generated with the procedure outlined above, we perform a likelihood analysis to determine the constraints on the VEP parameters. To~ensure consistency, we carefully follow the procedure adopted for the real IceCube data discussed in Ref.~\cite{Fiorillo:2020gsb}. We define the following chi-squared function:
\begin{equation}\label{eq:chi}
    \chi^2(\gamma_{21},\gamma_{31})=\text{min}_{\alpha,\beta}\left[\sum_i\frac{\left\{n_i-\alpha\left[1+\beta(0.5+\cos\theta_i)\right] \bar{n}^\text{VEP}_i(\gamma_{21},\gamma_{31})\right\}^2}{n_i+f_\mathrm{sys}^2 n_i^2} +\frac{(1-\alpha)^2}{\sigma_\alpha^2}+\frac{\beta^2}{\sigma_\beta^2}\right]\,,
\end{equation}
where $\sigma_\alpha=0.24$ and $\sigma_\beta=0.04$ quantify the uncertainty on the model of the atmospheric flux (see Refs.~\cite{Honda:2006qj,Esmaili:2014ota,Fiorillo:2020gsb}). The~quantities $\bar{n}^\text{VEP}_i(\gamma_{21},\gamma_{31})$ are the number of events expected in the VEP model, which depends on the VEP parameters. They are given by
\begin{equation}\label{eq:VEPexpected}
    \bar{n}^\text{VEP}_i(\gamma_{21},\gamma_{31}) = 2\pi T_\text{obs} \int \mathrm{d}E \int_{\cos\theta_i}^{\cos\theta_{i+1}} \mathrm{d}\cos\theta\, P_{\mu\mu}(E,\theta,\gamma_{21},\gamma_{31}) \frac{\mathrm{d}\Phi_\text{atm}}{\mathrm{d}E\mathrm{d}\Omega} A_\text{eff}(E)\,,
\end{equation}
where $P_{\mu\mu}$ is the probability of non-disappearance of muon neutrinos, obtained with the methods discussed in Section~\ref{sec:oscillations}. In~Equation~\eqref{eq:chi}, the~systematic error $f_\mathrm{rm}$ is considered as a free nuisance parameter. Hence, we assume that the systematic errors affecting the data are not known a priori. Following the procedure in Ref.~\cite{Fiorillo:2020gsb}, in~the data analysis the value of $f_\mathrm{sys}$ is chosen in such a way that the agreement with the data is within $2\sigma$ for the no-VEP~hypothesis.

For each mock data sample we determine the $90\%$ exclusion contour in the $\gamma_{21}\phi$--$\gamma_{31}\phi$ plane. We distinguish between two cases: $\gamma_{21}\times\gamma_{31}>0$ (direct ordering) and $\gamma_{21}\times\gamma_{31}<0$ (inverse ordering). These two cases have nothing to do with the direct and inverse ordering scheme of the neutrino masses, and~are chosen only as a nomenclature to distinguish them; the neutrino mass ordering is fixed to be direct in all the results obtained in this~work.

In Figure~\ref{fig:exclusioncontours10years}, we show the bands in which the exclusion contours lie for all Monte Carlo simulations. We take the observation time to be $T_\text{obs}=10~\mathrm{yr}$ and study the cases of direct and inverse ordering for the two benchmark choices $\bar{f}_\mathrm{sys}=1\%$ and $\bar{f}_\mathrm{sys}=4\%$ mentioned above. For~the two values of the systematic error, the~solid and dashed white lines represent the mean expected constrain. For~reference, we also show the previous constraints obtained with IceCube data~\cite{Fiorillo:2020gsb} (see also Ref.~\cite{Esmaili:2014ota}). Our results show that the KM3NeT ability of constraining the VEP model is similar to the one of IceCube. In~both telescopes, a~relevant role is played by the systematic uncertainty. Even for the case of $\bar{f}_\mathrm{sys}=1\%$, for~a number of events of the order of $10^5$ (typical value after 10 years of data taking), the~systematic error is much larger than the statistical one. This implies that an increase in the number of years, or~equivalently in the effective area, does not lead to significant improvements in the strength of the constraints. However, we emphasize that independent measurements from neutrino telescopes with different systematic errors other than IceCube are of paramount importance to place robust constraints on VEP parameters. On~the other hand, only a reduction in the systematic error of the experiments can lead to an improvement in the constraints. Decreasing the systematic error from $4\%$ to $1\%$ in the Monte Carlo generation does indeed lead to a narrowing of the band of expected exclusion contours and a strengthening of the~constraints.

%========================%
\section{Conclusions}\label{sec:concl}
%========================%
The equivalence principle is at the core of general relativity; as such, it provides a strong hint toward the construction of an ultraviolet completion of the gravitation theory. For~this reason, testing the equivalence principle might have profound implications for the future theory of gravity to be constructed. In~this work we have determined the sensitivity of the KM3NeT neutrino telescope to the effects of VEP on atmospheric neutrinos. Using the public KM3NeT effective area, we have simulated the angular distribution of the events collected over an observation time of 10 years. From~these simulated data, we have determined the 90\% allowed regions in the VEP parameter space for each data sample. Our study confirms that neutrino telescopes can indeed constrain VEP at a level of $\gamma\phi\sim 10^{-27}$, as~previous works have determined from the data observed by IceCube. These results are not directly comparable to the constraints obtained by experiments based on gravitational effects alone from VEP: the latter are in fact generally quantitatively weaker, but~do not stem from the assumption that VEP involves the neutrino sector. We find that, even with systematic errors of the order of $1\%$, 10 years of data are sufficient to enter the systematic-dominated regime. For~this reason, an~independent analysis with KM3NeT neutrino telescope, potentially affected by different systematic uncertainties than IceCube, is crucial to robustly investigate the equivalence~principle.

%%%%%%%%%%%%%%%%%%%%%%%%%%%%%%%%%%%%%%%%%%
\section*{Acknowledgments}
This work was supported by the research grant number 2017W4HA7S ``NAT-NET: Neutrino and Astroparticle Theory Network'' under the program PRIN 2017 funded by the Italian Ministero dell'Istruzione, dell'Universit\`a e della Ricerca (MIUR), and INFN Iniziativa Specifica TAsP.

%%%%%%%%%%%%%%%%%%%%%%%%%%%%%%%%%%%%%%%%%%

%=====================================

\end{document}